\documentclass[aps,prb,amsmath,amssymb,reprint,superscriptaddress,preprintnumbers,intlimits]{revtex4-1}
\usepackage{bm,latexsym,mathrsfs,enumerate}
\usepackage[mathcal]{euscript}
\usepackage[table,x11names]{xcolor}
\usepackage[breaklinks=true,
unicode=true,
urlcolor = RoyalBlue4,
colorlinks = true,
citecolor = RoyalBlue4,
linkcolor = RoyalBlue4,
pdftitle={ChiraMag},
pdfauthor={Denis Sheka}			
]{hyperref}
\usepackage{graphicx}
\usepackage{wrapfig}
\renewcommand{\vec}[1]{\bm{#1}}

\usepackage{savesym}
\usepackage[most]{tcolorbox}
\savesymbol{comment}
\usepackage[todonotes={textsize=tiny,textwidth=44pt},truncate=hyphenate,deletedmarkup=xout,addedmarkup=uwave]{changes}
\begin{document}
%\preprint{\textcolor[rgb]{0.00,0.50,0.75}{\texttt{Draft \gitAbbrevHash{} by \gitCommitterName{} on \gitCommitterIsoDate}}}
	
\title{The effect of curvature on the eigenexcitations of magnetic skyrmions}

\author{Anastasiia Korniienko}
\affiliation{Department of Physics, Technical University of Munich, 85748 Garching, Germany}
\affiliation{Helmholtz-Zentrum Dresden{\;-\;}Rossendorf e.V., Institute of Ion Beam Physics and Materials Research, 01328 Dresden, Germany}

\author{Attila K\'{a}kay}
\affiliation{Helmholtz-Zentrum Dresden{\;-\;}Rossendorf e.V., Institute of Ion Beam Physics and Materials Research, 01328 Dresden, Germany}

\author{Denis D. Sheka}
%\email{sheka@knu.ua}
\affiliation{Taras Shevchenko National University of Kyiv, 01601 Kyiv, Ukraine}

\author{Volodymyr P. Kravchuk}
\email{volodymyr.kravchuk@kit.edu}
\affiliation{Institut f\"ur Theoretische Festk\"orperphysik, Karlsruher Institut f\"ur Technologie, D-76131 Germany}
\affiliation{Bogolyubov Institute for Theoretical Physics of National Academy of Sciences of Ukraine, 03143 Kyiv, Ukraine}

\begin{abstract}
Spectrum of spin eigenmodes localized on a ferromagnetic skyrmion pinned by a geometrical defect (bump) of magnetic films is studied theoretically. By means of direct numerical solution of the corresponding eigenvalue problem and finite element micromagnetic simulations we demonstrate, that the curvature can induce localized modes with higher azimuthal and radial quantum numbers, which are absent for planar skyrmions (for the same parameters). The eigenfrequencies of all modes, except the breathing and gyromodes decreases with increasing curvature. Due to the translational symmetry break, the zero translational mode of the skyrmion gains a finite frequency and forms the gyromode, which describes the uniform rotation of skyrmions around the equilibrium position. In order to treat the gyromotion analytically we developed a Thiele-like collective variable approach. We show that N{\'e}el skyrmions in curvilinear films experience a driving force originating from the gradient of the mean curvature. The gyrofrequency of the pinned skyrmion is proportional to the second derivative of the mean curvature at the point of equilibrium. 
\end{abstract}

%\pacs{75.30.Et, 75.75.-c, 75.78.-n}
\maketitle

\section{Introduction}

Nowadays, magnetic skyrmions \cite{Bogdanov89r,Bogdanov94,Liu16a,Fert17,Wiesendanger16} are subject of extensive studies. This can be explained by the fundamental interest in the physics of topological solitons,\cite{Manton04} by the feasibility to use single skyrmions as possible bit carriers in various memory and logic devices, \cite{Fert13,Sampaio13,Tomasello14,Zhang15,Krause16,Kang16,Wiesendanger16,Mueller17,Fert17,Zhang15c} and also due to the ability to realize skyrmion lattices \cite{Muehlbauer09,Yu10,Yu11,Milde13,Roessler06} with possible applications in devices relying on the topological Hall effect.\cite{Lee09a,Neubauer09,Kanazawa11,Li13a} 

Introducing curvature by bending magnetic thin film the properties of magnetic skyrmions is changed. Majority of the new effects related to surface curvature can be explained in terms of the effective magnetic interactions caused by locally curved geometries: (i) curvilinear geometry-induced effective anisotropy and (ii) curvilinear geometry-induced effective Dzyaloshinskii-Moriya interaction (DMI).  \cite{Gaididei14,Sheka15,Streubel16a} This effective DMI emerges as an antisymmetric part of the common isotropic exchange in a curvilinear frame of reference which follows the surface. Transition into the curvilinear frame of reference is not just a mathematical trick, but is physically conditioned by the presence of the magnetic interactions determined by the film geometry, e.g. the uniaxial anisotropy whose axis follows the surface normal, or intrinsic DMI of the surface type.\cite{Crepieux98,Bogdanov01,Thiaville12} A convincing example of curvature induced DMI effect is the stabilization of ferromagnetic skyrmions on a curvilinear shell free of intrinsic DMI.\cite{Kravchuk16a,Pylypovskyi18a}

Previously it was reported \cite{Kravchuk18a} that a local curvature (bump) of the film can create attracting as well as repulsing potentials for skyrmions. In case of attraction the pinned skyrmion can posses a multiplet of states and the regular arrangement of the bumps will result in an artificial skyrmion lattice as the ground state of the system. \cite{Kravchuk18a} These interesting findings can be considered for applications. The present manuscript is a continuation of the static study presented in Ref.~\onlinecite{Kravchuk18a} and is focused on the linear dynamics of skyrmions pinned on bumps. 

Here, we report on the study of magnon eigenexcitations in ferromagnetic skyrmions pinned on geometrical defects with rotational symmetry. Three different methods are used, namely: (i) we formulate and solve numerically the eigenvalue problem for a skyrmion on a curvilinear bump. This is the generalization of the previously developed analysis in \cite{Kravchuk18} for the case of planar films. (ii) We perform full-scale finite element micromagnetic simulations using our code \text{TetraMag}.\cite{Kakay10} (iii) We generalize the Thiele equation for the case of a ferromagnetic topological soliton on a curvilinear film. The generalized equation keeps the form of common Thiele equation but the gyrovector follows the surface normal. 
%For the surfaces with nonzero Gau{\ss}ian curvature the absolute value of the gyrovector is generally not constant. However, 
The gyrovector amplitude approaches its planar value in the limit case when the skyrmion radius is much smaller as compared to the curvature radii. 
The curvature induced driving force proportional to the gradient of the mean curvature appears for the case of \emph{N{\'e}el} skyrmions. This effect takes place  due to the curvilinear geometry-induced effective DMI. It is analogous to the curvature induced domain walls motion in  curvilinear wires.\cite{Yershov18a,Yershov15b} We utilize the generalized Thiele equation in order to obtain the analytical expression for the gyromode of a small-radius N{\'e}el skyrmion. 

\section{Model}
The magnetic medium assumed to be a thin film of a chiral ferromagnet with perpendicular easy-axial anisotropy can be modeled by the following energy functional:

\begin{equation}\label{eq:E}
E=L\int\left[A \mathscr{E}_\text{x}+K(1-m_n^2)+\mathcal{D}\mathscr{E}_\textsc{d}\right]\mathrm{d}\mathcal{S},
\end{equation}
where $L$ is the film thickness and the integration is performed over the film area. The first term of the integrand is the exchange energy density with $\mathscr{E}_\text{x}=\sum_{i=x,y,z}(\partial_i\vec{m})^2$, and $A$ being the exchange constant. Here $\vec{m}=\vec{M}/M_s$ is the unit magnetization vector with $M_s$ being the saturation magnetization. The second term is the easy-normal anisotropy with $K>0$ and $m_n=\vec{m}\cdot\vec{n}$ is the normal magnetization component with $\vec{n}$ being the unit normal to the surface. The exchange-anisotropy competition results in the magnetic length $\ell=\sqrt{A/K}$, determining the characteristic length scale of the system (on which non-collinear magnetic structures can form). The last term in equation \eqref{eq:E} represents the Dzyaloshinskii-Moriya interaction with $\mathscr{E}_\textsc{d}=m_n\vec\nabla\cdot\vec{m}-\vec{m}\cdot\vec{\nabla}m_n$. Such kind of DMI originates from the inversion symmetry breaking at the film interface; it is typical for ultrathin films \cite{Crepieux98,Bogdanov01,Thiaville12} or bilayers, \cite{Yang15} and can result in formation of N\'{e}el (hedgehog) skyrmions.\cite{Sampaio13,Rohart13} The magnetostatic contribution is not included explicitly into our model, since in ultrathin films can be reduced as the renormalization of the anisotropy $K\to K^{\text{eff}}=K-2\pi M_s^2$,\cite{Carbou01,Fratta16b,Fratta19a} leading to an effective anisotropy constant. In equation \eqref{eq:E} it is assumed that the magnetization profile is uniform along the direction of the normal to the surface. The magnetization dynamics is described by the  Landau--Lifshitz--Gilbert equation
\begin{equation}\label{eq:LL}
\partial_t\vec{m}=\frac{\gamma_0}{M_s}\left[\vec{m}\times\frac{\delta E}{\delta\vec{m}}\right]+\eta\left[\vec{m}\times\partial_t\vec{m}\right],
\end{equation} 
where $\gamma_0$ is gyromagnetic ratio and $\eta$ is the Gilbert damping.

Our model system is a magnetic film with a curvilinear defect of rotational symmetry is considered. To this end we describe our film as a surface of revolution $\vec{\sigma}(s,\chi)=r(s)(\hat{\vec{x}}\cos\chi+\hat{\vec{y}}\sin\chi)+z(s)\hat{\vec{z}}$. Here $\chi\in[0,2\pi)$ is the azimuthal angle and $s\ge0$ is the radial distance along the surface. The pair of functions\footnote{These functions are interconnected by means of the relation $r'(s)^2+z'(s)^2\equiv1$. The requirement that the surface is smooth at $s=0$ imposes additional conditions on these functions: $z'(0)=0$, $r(0)=0$ and $r'(0)=1$. While the requirement of localization of the curvilinear defect results in $z(s)\to0$ and $r(s)\to s+\text{const}$, when $s\to\infty$. } $r(s)$ and $z(s)$ determine the shape of the surface as shown in Fig.~\ref{fig:profiles}(d): $r(s)$ and $z(s)$ denote the distance to the point of the surface with coordinate $s$ from the axis of revolution $\hat{\vec{z}}$ and from the $xy$-plane, respectively. The curvilinear properties of the surface $\vec{\sigma}$ are represented by the principal curvatures $k_1 = z''/r'$, $k_2=z'/r$. The prime denotes the derivatives with respect to $s$ throughout the whole manuscript. Note that $k_1(0)=k_2(0)=z''(0)$.

%\begin{figure}[h]
%	\includegraphics[width=0.45\textwidth]{rz.pdf}
%	\caption{The surface of revolution is made by means of rotation of the curve $\vec{\gamma}$ (green) around the vertical $z$-axis. Coordinate $s$ is the arc length of $\vec{\gamma}$ measured from the axis of rotation. Functions $z(s)$ and $r(s)$ are shown for the bump \eqref{eq:bump} with $r_0=1$.}\label{fig:zr}
%\end{figure}

%\nb[DS]{
%	Comments to Fig.~\ref{fig:profiles}:\\ (i) Yellow background for subfigures looks unusual;\\ (ii) The notation (a) has a wrong placement; \\
%	(iii) You compare profiles of planar skyrmion and skyrmion on a bump using: we already know \cite{Pylypovskyi18a} that such comparison should be done using surface polar radius \cite[see Eq.(2)]{Pylypovskyi18a}, but not the arc length.
%}[22.3.20]

\section{Static skyrmions}

\begin{figure}
	\includegraphics[width=0.9\columnwidth]{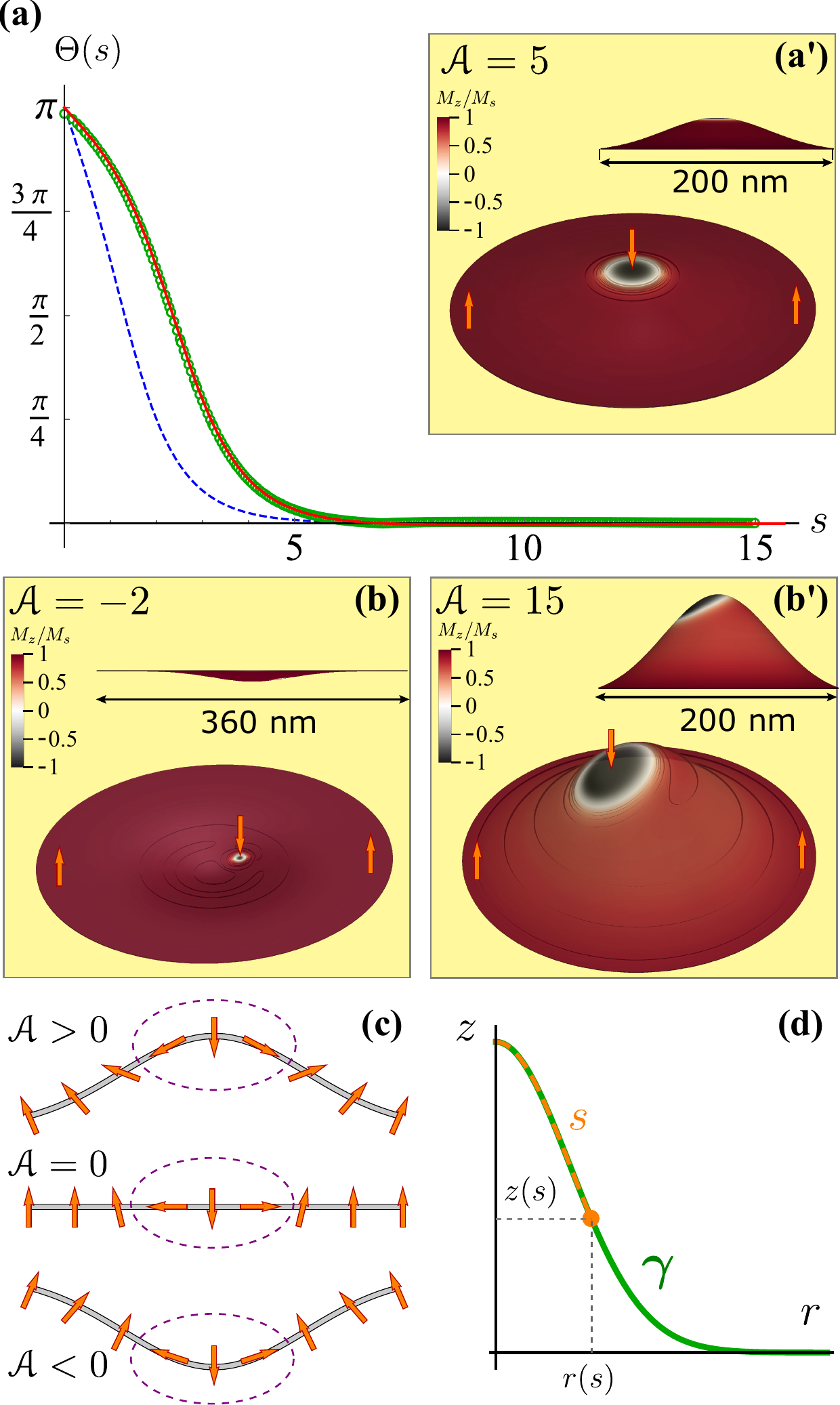}
	\caption{Skyrmion states of Gau{\ss}ian bumps \eqref{eq:bump} obtained for various bump amplitudes $\mathcal{A}=-2,5,15$ and fixed width $r_0=8$ and dimensionless DMI constant $d=0.98$. (a) -- stable skyrmion profile $\Theta(s)$ obtained for the bump with $\mathcal{A}=5$ by means of numerical solution of Eq.~\eqref{eq:Theta} (solid line) and micromagnetic simulations (green hollow dots) compared with the skyrmion profile on the planar film (dashed line). (a') The magnetisation profile obtained from micromagnetic simulations. (b) and (b') illustrate the cases when the bump generates a repulsive potential resulting in the skyrmion displacement from the central point. The origin of the pinning/repusive potentials is qualitatively sketched on inset (c): in comparison with the planar case ($\mathcal{A}$=0), the magnetization inside the dashed oval is more uniform (costs less exchange energy) and less uniform (costs more exchange energy) for the cases $\mathcal{A}>0$ and $\mathcal{A}<0$, respectively. Arrows in (a',b,b',c) show the magnetization. Panel (d) shows the bump profile, the geometrical definition of the parameter $s$ and functions $r(s)$, $z(s)$.}\label{fig:profiles}
\end{figure}

Utilizing the constraint $|\vec{m}|=1$ by means of the angular parameterization $\vec{m}=\sin\theta(\hat{\vec{s}}\cos\phi+\hat{\vec{\chi}}\sin\phi)+\hat{\vec{n}}\cos\theta$ with $\vec{n}=[\hat{\vec{s}}\times\hat{\vec{\chi}}]$ being the unit normal, one can show that Eq.~\eqref{eq:LL} has the static skyrmion solution in form of $\theta=\Theta(s)$ for $\phi=\Phi=0$. The skyrmion profile is determined by the following equation \cite{Kravchuk18a}
\begin{equation}\label{eq:Theta}
\nabla^2_s\Theta-\sin\Theta\cos\Theta\,\Xi+\frac{r'}{r}\sin^2\Theta(d-2k_2)=\mathcal{H}',
\end{equation}
where, $d=D/\sqrt{AK}$ is the dimensionless DMI constant, $\mathcal{H}=k_1+k_2$ is the mean curvature, and $\Xi=1+r^{-2}-2k_2^2+d\mathcal{H}$.
%\replaced[id=DS, remark = {\\ According to \cite{Kravchuk18a} we have another expression for $\Xi$}]{$\Xi = 1+r'^2/r^2-k_2^2+d H$ }{$\Xi=1+r^{-2}-2k_2^2+d\mathcal{H}$ }. 
In Eq.~\eqref{eq:Theta} and in the following all distances are measured in units of the magnetic length $\ell$. The radial part of the Laplace operator reads as $\nabla^2_sf=r^{-1}(rf')'$. Equation \eqref{eq:Theta} is solved with boundary conditions $\Theta(0)=\pi$, $\Theta(\infty)=0$. Note, that these boundary conditions correspond to the case of a relatively small curvature, when the multiplet of states are not allowed for the skyrmions.\cite{Kravchuk18a} Only singlet skyrmion states are therefore considered in this manuscript.

An example of a skyrmion profile generated by Eq.~\eqref{eq:Theta} for the case of a Gau{\ss}ian bump
\begin{equation}\label{eq:bump}
z=\mathcal{A}e^{-r^2/(2r_0^2)}
\end{equation}  
is shown in Fig.~\ref{fig:profiles}(a). As one may note, the skyrmion pinned on the bump has lager radius as compared to the planar case for the same intrinsic DMI and other material parameters. This effect can be interpreted as a result of an enhanced total DMI due to the curvature induced effective DMI.\cite{Gaididei14,Sheka15,Kravchuk16a} For the case $\mathcal{A}<0$ (concavity), the curvature induced DMI changes its sign decreasing the total DMI. This in turn decreases the skyrmion radius, see Fig.~\ref{fig:profiles}(b) for $\mathcal{A}=-2$. With the further increasing the negative amplitude of the concavity the intrinsic DMI is fully compensated by the curvature induced DMI and the skyrmion solution collapses to a point singularity. This effect was described previously for skyrmions on spherical shells.\cite{Kravchuk16a}

For a positive DMI constant, which is below the critical value $4/\pi$, a broad range of amplitudes $0<\mathcal{A}<\mathcal{A}_\text{max}(r_0,d)$ exists, for which the stable solution is a skyrmion centered on the bump (see Fig.~3 in Ref.~\onlinecite{Kravchuk18a}, which shows the diagram of skyrmion states on a Gaussian bump). For small negative $\mathcal{A}$ values, the center point of the bump becomes an unstable equilibrium position for the skyrmions. This instability effect originates from the exchange interaction, as qualitatively explained in Fig.~\ref{fig:profiles}(c). Namely, the spatial deformation of the film can reduce, as well as enlarge, the spatial gradients of the magnetization. 
%In curved geometries the energy degeneracy of the clockwise and counter-clockwise rotation of the spins is lifted, and moreover one of the rotations has such high energies that will be expelled from the region with gradient curvature. When keeping the rotation direction constant but changing the sign of the curvature has the same effect as changing the rotation direction, as qualitatively explained in Fig.~\ref{fig:profiles}(c), and as a result the skyrmion will move out from the middle of the bump.

Skyrmions can be again stabilised on the bump center for large negative $\mathcal{A}$ values. However, in this case several skyrmion solutions (multiplet) may appear, e.g. a doublet with small and large skyrmion radii, as reported previously in Ref.~\onlinecite{Kravchuk18a}. Alternatively, in order to stabilize skyrmions on the bump with negative amplitude (concave), one has to consider a negative DMI constant. This changes the skyrmion helicity to $\Phi=\pi$ and reverts the energies of the cases with $\mathcal{A}>0$ and $\mathcal{A}<0$ shown in Fig.~\ref{fig:profiles}(c). 

If a large-radius skyrmion is centered on the bump, the magnetization is not uniform in the skyrmion central area. This is because the magnetization tends to align to the normal direction due to the easy-normal anisotropy. This is in contrast to the planar case. With the increase of the amplitude $\mathcal{A}$, the magnetization nonuniformity increases resulting in an increase of the exchange energy. Finally, for $\mathcal{A}>\mathcal{A}_\text{max}$ the central equilibrium becomes unstable and the skyrmion shifts to the side of the bump, as shown in Fig.~\ref{fig:profiles}(b'). This is in line with the previous predictions. \cite{Kravchuk18a} So far from the Figs.~\ref{fig:profiles}(a',b,b') we can conclude that the theoretical predictions and the full-scale finite element micromagnetic simulations are in good agreement.

\section{Spectrum evaluation}
\label{sec:spectrum}

In the following let's consider an equilibrium skyrmion state centered on the bump. In order to study the spectrum of its linear excitations, we introduce small deviations in $\vartheta$ and $\varphi$:  $\theta=\Theta+\vartheta$, $\phi=\Phi+\varphi/\sin\Theta$. For the case of zero damping $\eta=0$, equation \eqref{eq:LL} can be linearized with respect to the excitations and results in 
\begin{equation}\label{eq:dev}
\begin{cases}
\partial_\tau\varphi=-\nabla^2\vartheta+U_1\vartheta+W\partial_\chi\varphi,\\
-\partial_\tau\vartheta=-\nabla^2\varphi+U_2\varphi-W\partial_\chi\vartheta.
\end{cases}
\end{equation}
The dimensionless time is introduced as $\tau=t\Omega_0$, where $\Omega_0=2\gamma_0K/M_s$. The Laplace operator has the form $\nabla^2=\nabla^2_s+r^{-2}\partial_{\chi\chi}^2$ and potentials have the following expressions:\cite{Kravchuk18a}
\begin{equation}\label{eq:UW}
\begin{split}
U_1=&\cos2\Theta\,\Xi-\frac{r'}{r}\sin2\Theta(d-2k_2),\\
U_2=&\cos^2\Theta\,\Xi-\Theta'^2+k_2^2-k_1^2-\Theta'(d-2k_1)\\
&-\frac{r'}{r}\sin\Theta\cos\Theta(d-2k_2),\\
W=&2\frac{r'}{r^2}\cos\Theta-\frac{1}{r}\sin\Theta(d-2k_2)
\end{split}
\end{equation}
The solutions of equations \eqref{eq:dev} are $\vartheta=f(s)\cos(\omega\tau+\mu\chi+\eta)$, $\varphi=g(s)\sin(\omega\tau+\mu\chi+\eta)$, where $\mu\in\mathbb{Z}$ is the azimuthal wave number and $\eta$ is an arbitrary phase. The eigenfrequencies $\omega$ and the corresponding eigenfunctions $f$, $g$ are determined by the following generalized eigenvalue problem (EVP)
\begin{equation}\label{eq:EVP}
\hat{H}\vec{\psi}=\omega\hat\sigma_1\vec\psi.
\end{equation}
Here $\vec{\psi}=(f,g)^\textsc{t}$, 
\begin{equation}\label{eq:H}
\hat{H}=\begin{pmatrix}
-\nabla^2_s+\frac{\mu^2}{r^2}+U_1&\mu W\\
\mu W&-\nabla^2_s+\frac{\mu^2}{r^2}+U_2
\end{pmatrix},
\end{equation}
and $\hat\sigma_1$ is the first Pauli matrix. In the limit case of a planar film ($k_1\equiv k_2\equiv 0$ and $r'\equiv1$) the formulated eigenvalue problem coincides with that formulated previously in Ref.~\onlinecite{Kravchuk18} for planar skyrmions.

\begin{figure*}
	\includegraphics[width=\textwidth]{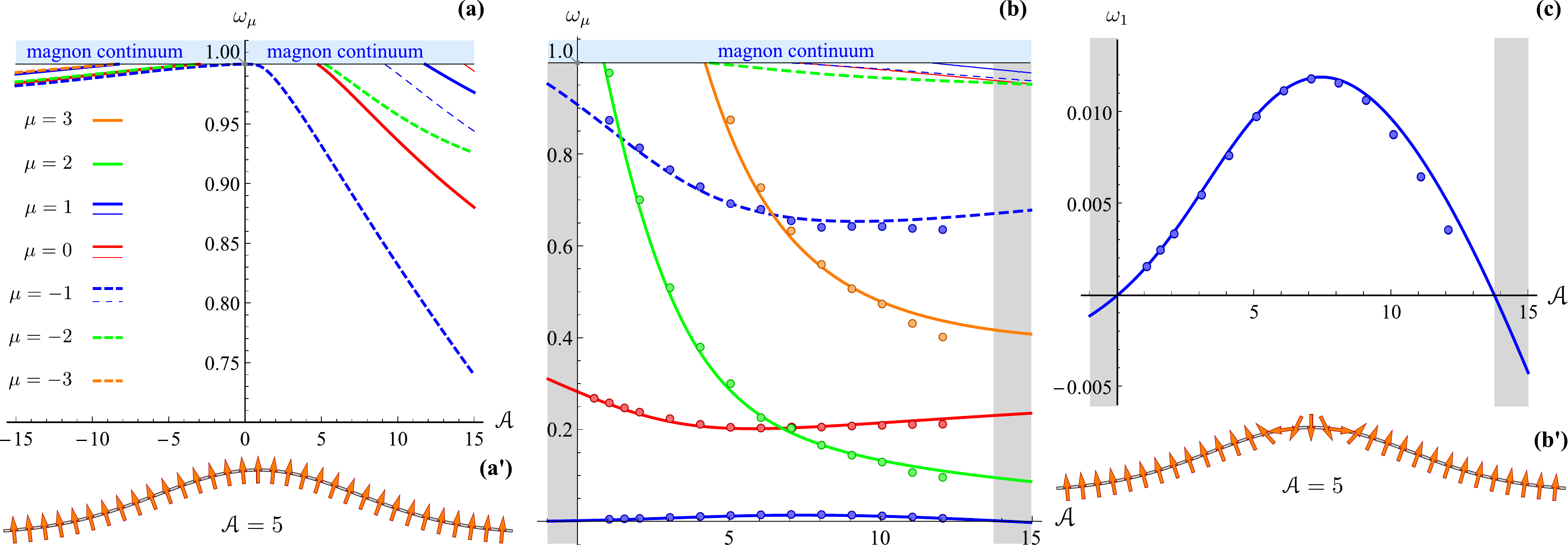}
	\caption{Eigenfrequencies of the localized eigenstates are obtained by means of numerical solution of EVP \eqref{eq:EVP} for the Gau{\ss}ian bump \eqref{eq:bump} with constant width $r_0=8$ and various amplitudes $\mathcal{A}$ are shown by lines. The markers show the eigenfrequencies obtained by means of micromagnetic simulations. The value of dimensionless DMI constant is $d=0.98$. The quasinormal and skyrmion equilibrium states obtained from Eq.~\eqref{eq:Theta} are shown on panels (a') and (b') for a particular bump height $\mathcal{A}=5$. The magnon excitation frequencies of the quasinormal and skyrmion states are summarized in (a) and (b) in function of the bump amplitude. Inset (c) is focused to the gyromode ($\mu=1$). The thin lines are eigenstate solutions whose functions $f$ and $g$ have a node along the radial coordinate $s$, see Fig.~\ref{fig:fg}. The gray shadowed rectangles denote the ranges of bump amplitudes ($\mathcal{A}<0$ and $\mathcal{A}>\mathcal{A}_{\text{max}}\approx13.77$) for which the bump center is not a stable equilibrium position of the skyrmion.}\label{fig:spectrum} 
\end{figure*}

An example of numerical solution of EVP is shown in Fig.~\ref{fig:spectrum}. In order to find out the influence of the curvature on the magnon spectrum, Gau{\ss}ian bumps with constant width $r_0$ and varying bump amplitudes $\mathcal{A}$ have been considered. Only spatially localized eigenstates are investigated. For the ground state, which is the quasinormal magnetization, a number of resonances located closely to the bottom edge of the magnon continuum appears, see Fig.~\ref{fig:spectrum}(a). In the planar thin-film limit, $\mathcal{A}\to0$, all localized states with $\mu\ne-1$ disappear. At the specific point of $\mathcal{A}=0$ the counterclockwise (CCW) mode $\mu=-1$ transforms to the Kittel mode with $\omega_{-1}=1$ and $f=g=\text{const}$. A close relation of the CCW localized mode and Kittel mode was previously indicated in Ref.~\onlinecite{Satywali18}.

The presence of the curvature significantly enriches the spectrum of the localized magnon states, see Fig.~\ref{fig:spectrum}(b). By means of the comparison with the previously studied skyrmion spectrum for a planar films, the following curvature induced effects can be distinguished. (i) First of all, it should be noted that due to the breaking of the translation symmetry the translational skyrmion mode ($\mu=1$) is transformed to the gyromode with small but nonzero frequency $\omega_{1}$, see Fig.~\ref{fig:spectrum}(c). The cases $\omega_{1}>0$ and  $\omega_{1}<0$ describe clockwise and counterclockwise skyrmion gyrations, respectively. The latter case corresponds to the repulsive effective potential or in other words -- to the unstable equilibrium position at the bump center. (ii) With increasing curvature higher modes with $\mu=\pm2$ (elliptic modes) and $\mu=3$ are generated. (iii) In contrast to the planar case the additional quantization of the localized modes in the radial direction is possible, see Fig.~\ref{fig:fg}. In Fig.~\ref{fig:spectrum}, the corresponding eigenfrequencies are shown by thin lines. (iv) Curvature leads to significant lowering of the CCW mode with $\mu=-1$, making feasible its experimental study.\cite{Satywali18}

\section{Collective variable approach for the gyromode}
Let us consider the low curvature limit $1\gg|\mathcal{H}|r_s\gg|\mathcal{K}|r_s^2$, where $\mathcal{K}$ is the Gau{\ss}ian curvature and $r_s$ is the skyrmion radius defined as $\Theta(r_s)=\pi/2$. In this particular case the motion of the skyrmion can be described using the rigid particle assumption, modeling the curvature as a perturbative potential, that has no influence on the skyrmion profile. In terms of the above introduced dimensionless units for the length and time, the corresponding Thiele equation (see Appendix~\ref{app:Thiele} for the derivation) reads as:
\begin{equation}\label{eq:Thiele}
\left[\vec{n}\times\partial_\tau\vec{R}\right]=\frac{\partial\mathcal{E}}{\partial\vec{R}}+\bar{\eta}\partial_\tau\vec{R},
\end{equation}
taking into account that the topological charge of the considered skyrmion on the plane is $N_{\text{top}}=-1$. Topological charge is a crucial quantity for the skyrmion dynamics in terms of collective coordinates, because it determines the amplitude of the gyrovector. One can show (see Appendix~\ref{app:Thiele}) that in the limit case of small curvature, the gyrovector amplitude approaches its planar value \eqref{eq:G-planar}. Thus, the planar topological charge can be used in this limit. Generalization of the skyrmion topological charge for an arbitrary curvature is an open question. In Ref.~\onlinecite{Kravchuk16a}, the degree $Q$ of the map $\sigma\mapsto S^2$ realized by the unit field $\vec{m}$ on $\sigma$ was proposed for such a generalization. Although $Q$ is an integer number which is invariant with respect to the continuous deformation of the magnetization, it is important to note that $Q$ isn't reflecting the dynamical properties of skyrmions in terms of the collective coordinates. This is reflected by Eq.~\eqref{eq:no}, see also the discussion after equation~\eqref{eq:G-general} in Appendix~\ref{app:Thiele}.

The total energy of the system, $\mathcal{E}=E/E_0$ is measured in units of $E_0=8\pi A L$, the position vector $\vec{R}=\vec{\sigma}(X^1,X^2)$ determines the skyrmion center, which has the curvilinear coordinates $X^1$ and $X^2$ on the surface. In the considered small curvature limit one can show that (see Appendix~\ref{app:Thiele})
\begin{equation}\label{eq:E-CV}
\mathcal{E}\approx\mathcal{C}\,\mathcal{H}(X^1,X^2)+\mathcal{E}_0,
\end{equation}
where energy $\mathcal{E}_0$ is independent on the collective coordinates $(X^1,X^2)$. Constant $\mathcal{C}$ is determined by the equilibrium skyrmion profile $\Theta_{\text{pl}}(s)$ for the case of a planar film, namely $\mathcal{C}=\mathcal{C}_1+\mathcal{C}_2d$, where constants $\mathcal{C}_1$ and $\mathcal{C}_2$ have form $\mathcal{C}_1=\frac{1}{4}\int_0^\infty\left[\Theta_{\text{pl}}(s)-\sin\Theta_{\text{pl}}(s)\cos\Theta_{\text{pl}}(s)\right]\mathrm{d}s$, $\mathcal{C}_2=\frac{1}{4}\int_0^\infty s\sin^2\Theta_{\text{pl}}(s)\mathrm{d}s$. The normalized damping constant $\bar\eta=\mathcal{C}_0\eta$ in \eqref{eq:Thiele} is also determined by the planar skyrmion profile: $\mathcal{C}_0=\frac{1}{4}\int_0^\infty\left[(\partial_s\Theta_{\text{pl}})^2+s^{-2}\sin^2\Theta_{\text{pl}}\right]s\,\mathrm{d}s$. Note that the energy expression \eqref{eq:E-CV} is correct for \emph{N{\'e}el} skyrmions only. For the case of a Bloch skyrmion one should take into account the quadratic terms in the curvature, which are neglected in the current study, see Appendix~\ref{app:thiele.skyrm} for details.
%\begin{equation}\label{eq:c0}
%c_0=\frac{1}{4}\int\limits_0^\infty\left[(\partial_s\Theta_{\text{pl}})^2+\frac{1}{s^2}\sin^2\Theta_{\text{pl}}\right]s\,\mathrm{d}s.
%\end{equation}

In order to handle equation \eqref{eq:Thiele}, one should note that $\partial_\tau\vec{R}=\vec{g}_\alpha\partial_\tau X^\alpha$ and $\frac{\partial \mathcal{E}}{\partial\vec{R}}=\vec{g}^\alpha\frac{\partial \mathcal{E}}{\partial X^\alpha}$,
%\begin{equation}\label{eq:Rt-E}
%\partial_\tau\vec{R}=\vec{g}_u\partial_\tau U+\vec{g}_v\partial_\tau V,\qquad \frac{\partial \mathcal{E}}{\partial\vec{R}}=\vec{g}^u\frac{\partial \mathcal{E}}{\partial U}+\vec{g}^v\frac{\partial \mathcal{E}}{\partial V},
%\end{equation}
where $\vec{g}_\alpha=\partial_\alpha\vec{\sigma}$, $\alpha\in\{1,2\}$ is the tangent curvilinear basis and $\vec{g}^\alpha$ is the corresponding dual basis.

Applying \eqref{eq:Thiele} and \eqref{eq:E-CV} for the surface of rotation $\vec{\sigma}=\sigma(s,\chi)$, one can show that the trajectory of motion of the skyrmion center, $s=s(\chi)$ is determined by the following relation:
\begin{equation}\label{eq:traj}
\int\limits_{s_0}^{s(\chi)}\frac{\mathrm{d}s'}{r(s')}=\bar\eta(\chi-\chi_0),
\end{equation}
where $(s_0,\chi_0)$ is the initial skyrmion position. In the vicinity of the central point $s=0$ the trajectory \eqref{eq:traj} is approximated by a spiral $s(\chi)\approx s_0e^{\bar{\eta}(\chi-\chi_0)}$. The velocity of the skyrmion therefore is:
\begin{equation}\label{eq:v}
\vec{v}=-\frac{\mathcal{C}\mathcal{H}'(s)}{1+\bar{\eta}^2}\left[\bar{\eta}\frac{r'(s)}{r(s)}\hat{\vec{s}}+\hat{\vec{\chi}}\right].
\end{equation}
As it follows from \eqref{eq:v}, the skyrmion is immobile on a surface with constant mean curvature (e.g. surface of a sphere, minimal surface). The central point $s=0$ is a stationary point which is stable (unstable) if $\mathcal{H}''(0)>0$ ($\mathcal{H}''(0)<0$). This result supports the qualitative explanations of the skyrmion stability shown in Fig.~\ref{fig:profiles}(c).

\begin{figure}
	\includegraphics[width=\columnwidth]{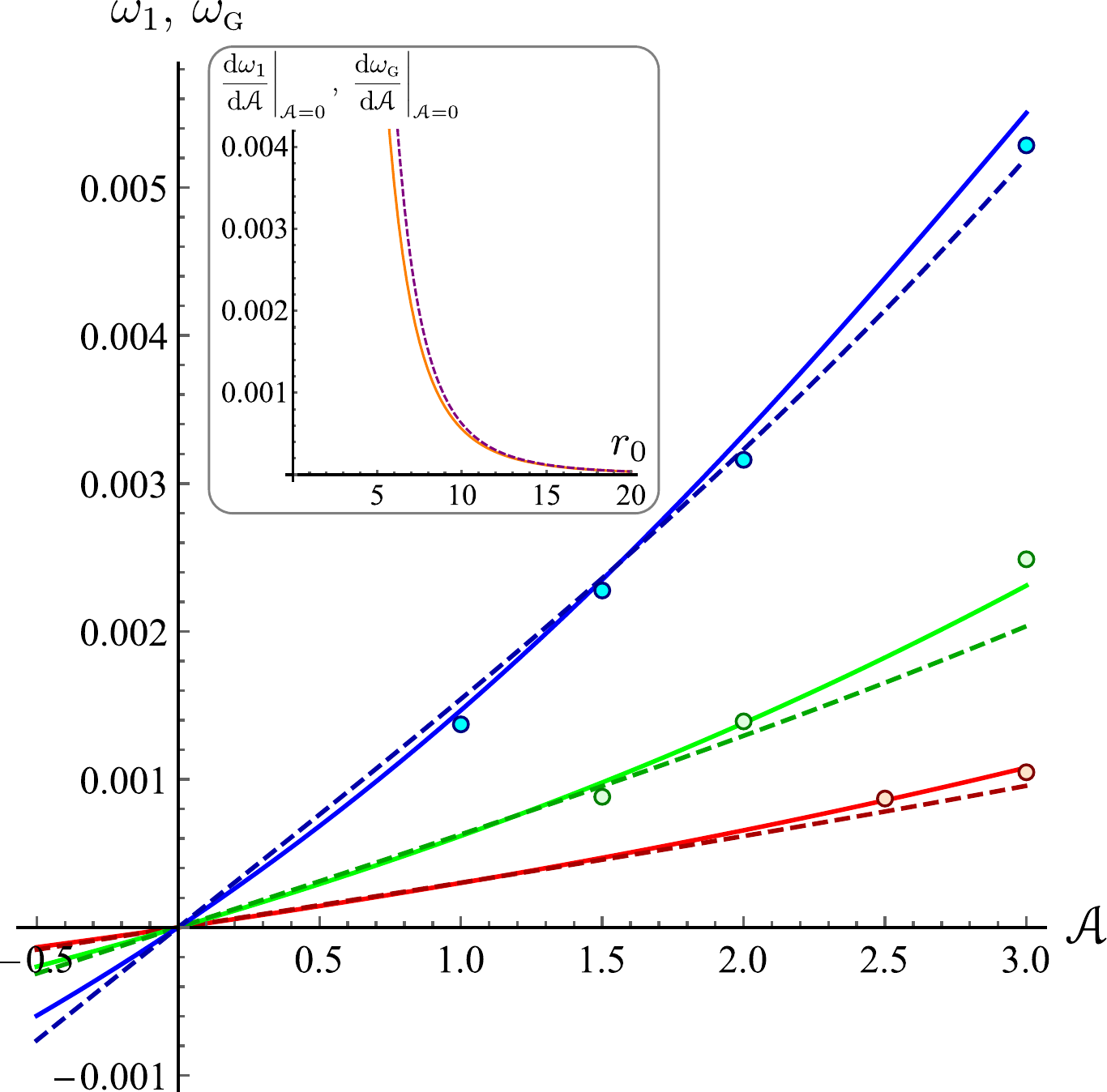}
	\caption{Frequency of the skyrmion gyromotion in the vicinity of the center of the low-amplitudes bumps \eqref{eq:bump} for three different bump radii are obtained in three different ways: (i) by means of numerical solution of EVP \eqref{eq:EVP} for $\mu=1$ (solid lines), (ii) by means of the collective variable approximation \eqref{eq:omega1} (dashed lines), and (iii) by means of micromagnetic simulations (markers). For all cases $d=0.98$ which results in $\mathcal{C}\approx1.556$. The inset demonstrates the slope in the point $\mathcal{A}=0$, the dashed line corresponds to the approximation $\mathrm{d}\omega_{\textsc{g}}/\mathrm{d}\mathcal{A}|_{\mathcal{A}=0}\approx4\mathcal{C}/r_0^4$.
	}\label{fig:gyro}
\end{figure}

For zero damping the skyrmion rotates around the stationary point with constant frequency $\omega_\textsc{g}=\mathcal{C}\mathcal{H}'(s_0)/r(s_0)$
which is determined by the initial skyrmion displacement. Here the positive frequency sign corresponds to the clockwise rotation. In the limiting case of  infinitesimal displacements, $s_0\to0$, the skyrmion gyration is described by the magnon mode with $\mu=1$, which was discussed in the previous section. In this case
\begin{equation}\label{eq:omega1}
\omega_1=\omega_\textsc{g}\approx\mathcal{C}\mathcal{H}''(0).
\end{equation}
For the Gau{\ss}ian bump \eqref{eq:bump} one has
$\mathcal{H}''(0)=4\frac{\mathcal{A}}{r_0^4}(1 + \frac{\mathcal{A}^2}{r_0^2})$. 
The corresponding comparison of the collective variable approach predictions in Eq. \eqref{eq:omega1} for $\omega_\textsc{g}$, the numerical calculations for $\omega_1$ and the gyromode frequencies from micromagnetic simulations for three different bump radii are shown in Fig.~\ref{fig:gyro}. Please note the good agreement between the collective coordinates and the EVP solutions. The small deviation of the slopes $\mathrm{d}\omega_{\textsc{g}}/\mathrm{d}\mathcal{A}|_{\mathcal{A}=0}$ and $\mathrm{d}\omega_{1}/\mathrm{d}\mathcal{A}|_{\mathcal{A}=0}$, which appear for small $r_0$ can be explained by the fact that the energy approximation \eqref{eq:E-CV}, as well as the  magnitude of the gyrovector \eqref{eq:G-planar} used in \eqref{eq:Thiele}, were obtained under the assumption that the metric is constant within the skyrmion area. This approximation is violated if the skyrmion radius $r_s$ is comparable with the bump width $r_0$.

\section{Conclusions}
	 We presented an analysis on the influence of the curvature on the spectrum of localized magnon eigenmodes of ferromagnetic skyrmions. We demonstrated that the curvature induces modes with higher azimuthal quantum numbers, which are absent for skyrmions in planar films. Additionally, modes with radial quantum numbers can appear for the skyrmions pinned on a bump. 
	 
	 Interestingly, the translational mode of the skyrmion is transformed into the gyromode by the curvature and has nonzero frequency proportional to the second derivative of the mean curvature at the bump center $\omega_{\textsc{g}}\propto\mathcal{H}''(0)$. For small amplitude Gau{\ss}ian bumbs \eqref{eq:bump} the following simple relation is found: $\omega_{\textsc{g}}\propto\mathcal{A}/r_0^4$. 
	 We have shown that these analytical estimations can be obtained with the Thiele equation, which we generalized for the case of skyrmions on an arbitrary curvilinear shell.  Furthermore, we demonstrated that N{\'e}el skyrmions experience the curvature induced driving force proportional to the gradient of the mean curvature.
	 	
\section{Acknowledgments}
We thank Denys Makarov for the fruitful discussions. A.\,Korniienko acknowledges financial support from DAAD (Leonhard Euler Programm, Projekt-ID: 57430566). A.~K\'{a}kay acknowledges the financial support of within the DFG programme KA 5069/1-1. D.\,Sheka acknowledges the financial support from the Alexander von Humboldt Foundation (Research Group Linkage Programme). In part, this work was supported by the Program of Fundamental Research of the Department of Physics and Astronomy of the National Academy of Sciences of Ukraine (Project No. 0116U003192), by Taras Shevchenko National University of Kyiv (Project No. 19BF052-01).  	
\appendix

\section{Eigenfunctions}
\begin{figure*}
	\includegraphics[width=\textwidth]{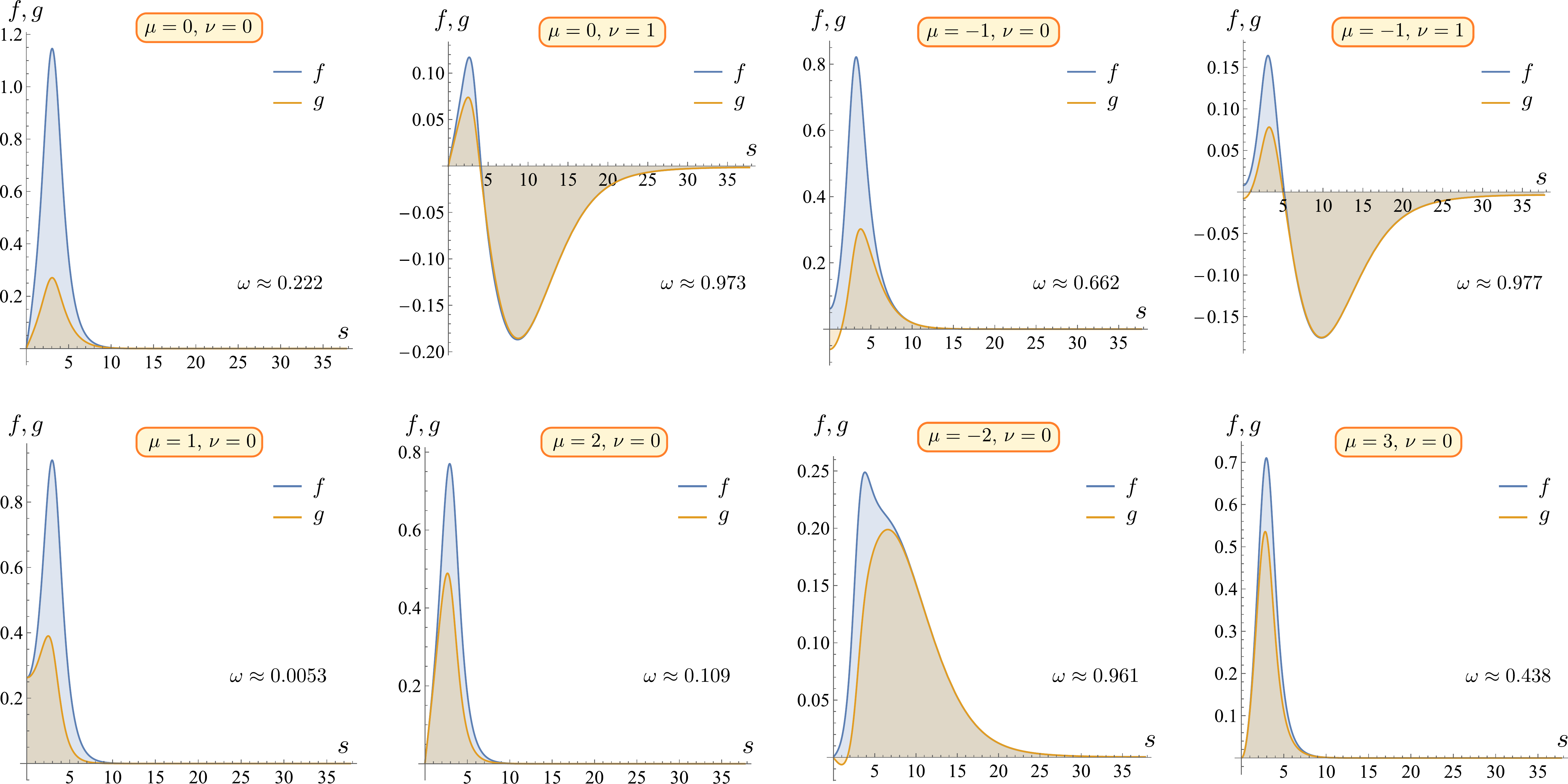}
	\caption{Eigenfunctions of the localized modes shown in Fig.~\ref{fig:spectrum}(b) for the bump amplitude $\mathcal{A}=12$. $\mu$ and $\nu$ denote the azimuthal and radial quantum numbers, respectively.}\label{fig:fg}
\end{figure*}

Some examples of eigenfunctions $f(s)$ and $g(s)$ of localized eigenmodes are shown in Fig.~\ref{fig:fg}. In contrast to the localized modes of the planar skyrmion, the additional quantization in radial direction ($\nu>0$) takes place for the skyrmion on a bump. 

Eigenfunctions $f$ and $g$ satisfy orthogonality condition (for given $\mu$)
\begin{equation}\label{eq:orth}
\int_{0}^{\infty}\left[f_{\mu,\nu}(s)g_{\mu,\nu'}(s)+f_{\mu,\nu'}(s)g_{\mu,\nu}(s)\right]r(s)\mathrm{d}s=\delta_{\nu,\nu'}
\end{equation}
which is similar as for the case of a planar skyrmion.\cite{Kravchuk18}
%Normalization of the eigenfunctions in Fig.~\ref{fig:fg} corresponds to \eqref{eq:orth}, namely $\int_{0}^{\infty}f_{\mu,\nu}(s)g_{\mu,\nu'}(s)r(s)\mathrm{d}s=1/2$.

\section{Curvilinear generalization of Thiele equation}\label{app:Thiele}
\subsection{General 3D case}
Assume that magnetization $\vec{m}(\vec{r},t)$ in some 3D space domain $\vec{\mathfrak{r}}$ can be presented in form $\vec{m}=\vec{m}(\vec{r},X^1(t),X^2(t),\dots)$ where $X^i(t)$ are some collective variables. Multiplying Landau-Lifshitz  equation \eqref{eq:LL} first by $\partial\vec{m}/\partial X^i\times(\dots)$ then by $\vec{m}\cdot(\dots)$ and then integrating over the space domain $\vec{\mathfrak{r}}$ one obtains the well known \cite{Mertens00,Tretiakov08} collective variable equation
\begin{equation}\label{eq:col-var-tensor}
\mathbb{G}_{ij}\partial_tX^{j}=\frac{\partial E}{\partial X^i}+\eta\mathbb{D}_{ij}\partial_tX^{j},
\end{equation}
where
\begin{subequations}\label{eq:G-D-tensors}
	\begin{align}
	\label{eq:G-tensor}
&\mathbb{G}_{ij}=\frac{M_s}{\gamma_0}\int_{\vec{\mathfrak{r}}}\vec{m}\cdot\left(\frac{\partial\vec{m}}{\partial X^i}\times\frac{\partial\vec{m}}{\partial X^j}\right)\mathrm{d}V,\\
\label{eq:D-tensor}
&\mathbb{D}_{ij}=\frac{M_s}{\gamma_0}\int_{\vec{\mathfrak{r}}}\frac{\partial\vec{m}}{\partial X^i}\cdot\frac{\partial\vec{m}}{\partial X^j}\mathrm{d}V
\end{align}
\end{subequations}
with $\mathrm{d}V$ being the volume element. Let us assume now that there exists a curvilinear frame of reference $\{\xi^1,\xi^2,\xi^3\}$ in $\vec{\mathfrak{r}}$ such that the magnetization dynamics can be presented in the form of a traveling wave 
\begin{equation}\label{eq:travel-wave}
m^i=m^i(\xi^1-X^1,\xi^2-X^2,\xi^3-X^3).
\end{equation}
 Here $\vec{m}=m^i\tilde{\vec{g}}_i$, where $\tilde{\vec{g}}_i=\partial_i\vec{\mathfrak{r}}$ is covariant (tangent) basis \footnote{Vectors $\tilde{\vec{g}}_i$ compose a basis in the tangent space.} induced by the parameterization $\vec{\mathfrak{r}}=\vec{\mathfrak{r}}(\xi^1,\xi^2,\xi^3)$. The shortening $\partial_i=\partial_{\xi^i}$ is used here and below. Note that the traveling-wave model formulated for the Cartesian coordinates and for Cartesian magnetization components describes only translations in 3D space and do not describe possible soliton rotations. The latter can be taken into account by introducing a curvilinear frame of reference whose local basis rotates together with the soliton. 

Since the equation \eqref{eq:col-var-tensor} is derived in coordinate-independent way, it keeps its form for any frame of reference. The tensors \eqref{eq:G-D-tensors} now read
\begin{subequations}\label{eq:G-D-tensors-crv}
	\begin{align}
	\label{eq:G-tensor-crv}
	&\mathbb{G}_{ij}=\frac{M_s}{\gamma_0}\int_{\vec{\mathfrak{r}}}\varepsilon_{kln}m^k\partial_i m^l\partial_j m^n\mathrm{d}V,\\
	\label{eq:D-tensor-crv}
	&\mathbb{D}_{ij}=\frac{M_s}{\gamma_0}\int_{\vec{\mathfrak{r}}}\tilde{g}_{kl}\partial_im^k\partial_jm^l\mathrm{d}V.
	\end{align}
\end{subequations}
Here $\tilde{g}_{kl}=\tilde{\vec{g}}_k\cdot\tilde{\vec{g}}_l$ is the metric tensor and $\varepsilon_{kln}=\sqrt{|\tilde{g}|}\epsilon_{kln}$ is Levi-Civita tensor, with $\tilde{g}=\det||\tilde{g}_{ij}||$ and $\epsilon_{kln}$ being the Levi-Civita symbol. We also took into account that $\partial_{X^i}m^k=-\partial_{i}m^k$. It is important to note that derivatives in \eqref{eq:G-D-tensors-crv} are not covariant and for this reason generally
\begin{equation}\label{eq:no}
\varepsilon_{kln}m^k\partial_i m^l\partial_j m^n \ne \vec{m}\cdot[\partial_i\vec{m}\times\partial_j\vec{m}].
\end{equation}
The equality in \eqref{eq:no} takes place only for Euclidian metric when $\tilde{g}_{kl}$ are coordinate independent, e.g. for Cartesian frame of reference.

Since tensor $\mathbb{G}_{ij}$ is by definition antisymmetric, one can write $\mathbb{G}_{ij}=\varepsilon_{ijk}G^k$. By means of the gyrovector $\vec{G}=\tilde{\vec{g}}_kG^k$ one can rewrite \eqref{eq:col-var-tensor} in a common Thiele form
\begin{equation}\label{eq:Thiele-crv}
\left[\partial_t\vec{R}\times\vec{G}\right]=\frac{\partial E}{\partial\vec{R}}+\eta\mathbf{D}\partial_t\vec{R}.
\end{equation} 
Here we introduced the notation $\tilde{\vec{g}}_i\partial_tX^i=\partial_t \vec{R}$, which defines $\vec{R}$ as a vector of the soliton position
\begin{equation}\label{eq:R-pos}
\vec{R}=\vec{\mathfrak{r}}(X^1,X^2,X^3).
\end{equation}
We also introduced the notation 
\begin{equation}\label{eq:dEdR}
\frac{\partial E}{\partial\vec{R}}=\tilde{\vec{g}}^i\frac{\partial E}{\partial{X^i}},
\end{equation}
which is consistent with \eqref{eq:R-pos}, here $\tilde{\vec{g}}^i$ are vectors of the dual basis. Indeed, using \eqref{eq:R-pos} one can write $\partial E/\partial X^i=(\partial E/\partial\vec{R})\cdot(\partial\vec{R}/\partial X^i)=(\partial E/\partial\vec{R})\cdot\tilde{\vec{g}}_i$. Multiplying this equation by $\tilde{\vec{g}}^i$ and using the identity $\vec{a}=\tilde{\vec{g}}^i(\vec{a}\cdot\tilde{\vec{g}}_i)$ one obtains \eqref{eq:dEdR}. The damping tensor is $\mathbf{D}=\mathbb{D}_{ij}\tilde{\vec{g}}^i\otimes\tilde{\vec{g}}^j$.\footnote{Note that $(\vec{a}\otimes\vec{b})\vec{c}=\vec{a}(\vec{b}\cdot\vec{c})$} Components of the gyrovector can be expressed as follows~\footnote{One should use here the property $\varepsilon_{ikl}\varepsilon^{jkl}=2\delta_i^j$.}
\begin{equation}\label{eq:G}
G^i=\frac12\varepsilon^{ijk}\mathbb{G}_{jk}.
\end{equation}
Note that all quantities in this equation generally depend on $X^1,X^2,X^3$.
%or in the explicit form
%\begin{equation}\label{eq:G-vector}
%G^k=\frac{M_s}{\gamma_0}\frac{\varepsilon^{ijk}(X^1,X^2,X^3)}{2}\int_{\vec{\mathfrak{r}}}\varepsilon_{kln}(\xi^1,\xi^2,\xi^3)m^k\partial_i m^l\partial_j m^n\mathrm{d}V
%\end{equation}
\subsection{Thin curvilinear shell}
Let us consider the space domain in form of curvilinear shell of thickness $L$
\begin{equation}\label{eq:shell}
\vec{\mathfrak{r}}(\xi^1,\xi^2,\xi^3)=\vec{\sigma}(\xi^1,\xi^2)+\vec{n}(\xi^1,\xi^2)\xi^3,
\end{equation}
 where $\vec{\sigma}(\xi^1,\xi^2)$ is central shell surface, $\xi^3\in[-L/2,L/2]$ and $\vec{n}=\vec{g}_1\times\vec{g}_2/\sqrt{|g|}$ is unit normal to the surface. Here $\vec{g}_\alpha=\partial_\alpha\vec{\sigma}=\lim_{\xi^3\to0}\tilde{\vec{g}}_\alpha$ is the covariant (tangent) basis on the surface. Here and below Greek indices take values $\{1,2\}$, while Latin indices take the values $\{1,2,3\}$. $g=\det||g_{\alpha\beta}||$, with $g_{\alpha\beta}=\vec{g}_\alpha\cdot\vec{g}_\beta$ being the surface metric tensor.
 
Let magnetization in $\vec{\mathfrak{r}}$ satisfies the condition \eqref{eq:travel-wave} and does not depend on $\xi^3$. Consequently, the total energy $E$ does not depend on $X^3$. In this case one can easily show that Thiele equation \eqref{eq:Thiele-crv} is fulfilled if $X^3\equiv0$.\footnote{Strictly speaking $X^3\equiv\text{const}$ The choice $X^3\equiv0$ is convenient because in this case one can work only with the metric on the surface $\vec{\sigma}$.}, i.e.  
\begin{equation}\label{eq:R}
\vec{R}=\vec{\sigma}(X^1,X^2).
\end{equation}
%and $X^3=0$. Indeed, in this case $\partial_t\vec{R}=\partial_{X^\alpha}\vec{\sigma}\partial_tX^\alpha=\vec{g}_\alpha\partial_tX^\alpha$. \highlight[id=DS, remark = {That was not a definition, just an expression.}]{This coincides with the definition} for $\partial_t\vec{R}$ used in \eqref{eq:Thiele-crv} because $\partial_tX^3=0$. For the total energy $E$ one can write $\partial E/\partial X^\alpha=(\partial E/\partial\vec{R})\cdot(\partial\vec{R}/\partial X^\alpha)=(\partial E/\partial\vec{R})\cdot\vec{g}_\alpha$. Multiplying this equation by $\vec{g}^\alpha$ and using the identity $\vec{a}=\vec{g}^\alpha(\vec{a}\cdot\vec{g}_\alpha)$ one obtains the relation $\partial E/\partial\vec{R}=\vec{g}^\alpha\partial E/\partial X^\alpha$ which  \highlight[id=DS, remark = {I think that was not a definition, just an expression.}]{coincides with the corresponding definition} for $\partial E/\partial\vec{R}$ used in \eqref{eq:Thiele-crv} if $E$ does not depend on $X^3$ and $X^3=0$.

%\begin{widetext}
Since magnetization does not depend on $\xi^3$, according to \eqref{eq:G} and \eqref{eq:G-tensor-crv} one has $\vec{G}=G^3\tilde{\vec{g}}_3=G\vec{n}$, where $G=G^3$ is amplitude of the gyrovector:
\begin{equation}\label{eq:G-general}
\begin{aligned}
&G=\frac{M_s}{\gamma_0}\frac{L}{\sqrt{|g(X^1,X^2)|}}\\
&\int_{\vec{\sigma}}\sqrt{|g(\xi^1,\xi^2)|}\begin{vmatrix}
	m^1 & m^2 & m^3 \\
	\partial_1m^1 & \partial_1m^2 & \partial_1m^3 \\
	\partial_2m^1 & \partial_2m^2 & \partial_2m^3 
	\end{vmatrix}\mathrm{d}S\\
	&+(\text{higher terms in }L),
\end{aligned}
\end{equation}
%\comment[id=DS]{
%I suggest to present \eqref{eq:G-general} in a bit compact form and explain that we consider the linear part in thickness, also to write explicitly that $\varepsilon_{kln} = \epsilon_{kln}$ $\sqrt{|g(\xi^1,\xi^2)|}$.}
%\begin{equation}\label{eq:G-general-1}
%\added[id=DS]{
%G =\frac{M_S}{\gamma_0}\frac{L}{\sqrt{|g(X^1,X^2)|}} \!\! \int_{\vec{\sigma}}\!\! \varepsilon_{kln} m^k\partial_1 m^l \partial_2 m^n \mathrm{d}S
%}
%\end{equation}
where $\mathrm{d}S=\sqrt{|g|}\mathrm{d}\xi^1\mathrm{d}\xi^2$ is element of the surface area. For a constant magnetization (in the curvilinear frame of reference), the integrand in \eqref{eq:G-general} vanishes as well as the left-hand side of \eqref{eq:no}. However, the right-hand side of \eqref{eq:no} generally does not vanish, e.g. it is the Gau{\ss}ian curvature for the normal magnetization ($m^1=m^2=0$, $m^3=1$).
%\end{widetext}
The linear in $L$ part of the dissipative tensor is $\mathbf{D}=D_{\alpha\beta}\vec{g}^\alpha\otimes\vec{g}^\beta$, where
\begin{equation}\label{eq:Dab}
D_{\alpha\beta}=\frac{M_s}{\gamma_0}L\int_{\vec{\sigma}}\left({g}_{\mu\nu}\partial_\alpha m^\mu\partial_\beta m^\nu+\partial_\alpha m^3\partial_\beta m^3\right)\mathrm{d}S.
\end{equation}
Here we also utilized the independence of magnetization on $\xi^3$.

If the basis vectors $\vec{g}_1$ and $\vec{g}_2$ are orthogonal then one can introduce the angular parameterization for the magnetization: $\sqrt{g_{11}}m^1=\sin\theta\cos\phi$, $\sqrt{g_{22}}m^2=\sin\theta\sin\phi$ and $m^3=\cos\theta$. In this case one obtains 
%\begin{equation}\label{eq:det}
%\begin{split}
%&\begin{vmatrix}
%m^1 & m^2 & m^3 \\
%\partial_1m^1 & \partial_1m^2 & \partial_1m^3 \\
%\partial_2m^1 & \partial_2m^2 & \partial_2m^3 
%\end{vmatrix}=\sin\theta\left[\vec{\nabla}\theta\times\vec{\nabla}\phi\right]\cdot\vec{n}\\
%&+\sin\theta\sin\phi\cos\phi\left[\vec{\nabla}\theta\times\vec{\nabla}\ln\sqrt{\frac{g_{11}}{g_{22}}}\right]\cdot\vec{n}\\
%&+\sin^2\theta\cos\theta[\vec{\nabla}\phi\times(\cos^2\phi\vec{\nabla}\ln\sqrt{g_{11}}\\
%&-\sin^2\phi\vec{\nabla}\ln\sqrt{g_{22}})]\cdot\vec{n}\\
%&+\sin^2\theta\cos\theta\sin\phi\cos\phi\left[\vec{\nabla}\ln\sqrt{g_{11}}\times\vec{\nabla}\ln\sqrt{g_{22}}\right]\cdot\vec{n},
%\end{split}
%\end{equation}
\begin{subequations}\label{eq:G-explicit}
\begin{equation}\label{eq:G-int}
G=\frac{M_s}{\gamma_0}\frac{L}{\sqrt{|g(X^1,X^2)|}} \int_{\vec{\sigma}}\!\sqrt{|g(\xi^1,\xi^2)|}\;\vec{\mathcal{G}} \cdot \vec{\mathrm{d}}\vec{S}
\end{equation}
\begin{equation}\label{eq:G-den}
\begin{aligned}
\vec{\mathcal{G}}&=\sin\theta\left[\vec{\nabla}\theta\times\vec{\nabla}\phi\right]\\
&+\sin\theta\sin\phi\cos\phi\left[\vec{\nabla}\theta\times\vec{\nabla}\ln\sqrt{\frac{g_{11}}{g_{22}}}\right]\\
&+\sin^2\theta\cos\theta[\vec{\nabla}\phi\times(\cos^2\phi\vec{\nabla}\ln\sqrt{g_{11}}\\
&+ \sin^2\phi\vec{\nabla}\ln\sqrt{g_{22}})]\\
&+\sin^2\!\theta\cos\theta\sin\phi\cos\phi \! \left[\vec{\nabla}\ln\! \sqrt{g_{11}}\!\times\!\vec{\nabla}\ln\! \sqrt{g_{22}}\right]\!,
\!\!\!\!\!\!\!\!\!\!\!\!\!\!\!\!\!\!\!\!
\end{aligned}
\end{equation}
\end{subequations}
where $\vec{\nabla}=\vec{g}^\alpha\partial_\alpha$ is the surface del operator and $\vec{\mathrm{d}}\vec{S}=\vec{n}\mathrm{d}S$. 
%\nb[DS]{
%I got instead of \eqref{eq:det} another equation, see \eqref{eq:G-explicit}.\\
%(i) The LHS in \eqref{eq:det} should have extra $\sqrt{|g|}$ in front of determinant. Otherwise \eqref{eq:G-simple} contradicts to \eqref{eq:det}.\\
%(ii) I got additional factor $1/\sqrt{|g|}$ in the RHS of \eqref{eq:det}, however it cancels with $\mathrm{d}S = \sqrt{|g|}\mathrm{d}\xi^1 \mathrm{d}\xi^2$.\\
%(iii) I got an opposite sign in the 4th line.\\
%(iv) I suggest to write down instead of \eqref{eq:det} the explicit form of $G$ as follows:
%\begin{equation}\label{eq:G-explicit}
%\begin{aligned}
%G &=\frac{M_S}{\gamma_0}\frac{L}{\sqrt{|g(X^1,X^2)|}} \!\! \int_{\vec{\sigma}}\!\! \left(\vec{\mathcal{J}} \cdot \vec{n} \right) \mathrm{d}\xi^1 \mathrm{d}\xi^2\\
%\vec{\mathcal{J}}&=\sin\theta\left[\vec{\nabla}\theta\times\vec{\nabla}\phi\right]\\
%&+\sin\theta\sin\phi\cos\phi\left[\vec{\nabla}\theta\times\vec{\nabla}\ln\sqrt{\frac{g_{11}}{g_{22}}}\right]\\
%&+\sin^2\theta\cos\theta[\vec{\nabla}\phi\times(\cos^2\phi\vec{\nabla}\ln\sqrt{g_{11}}\\
%&\alert{+} \sin^2\phi\vec{\nabla}\ln\sqrt{g_{22}})]\\
%&+\sin^2\!\theta\cos\theta\sin\phi\cos\phi \! \left[\vec{\nabla}\ln\! \sqrt{g_{11}}\!\times\!\vec{\nabla}\ln\! \sqrt{g_{22}}\right]\!,
%\!\!\!\!\!\!\!\!\!\!\!\!\!\!\!\!\!\!\!\!
%\end{aligned}
%\end{equation}
%}[20.3.20]
For the case of Euclidean (coordinate independent) metric the expression \eqref{eq:G-explicit}  transforms into the familiar \cite{Malozemoff79,Huber82a} formula
\begin{equation}\label{eq:G-simple}
G^\textsc{e}=\frac{M_s}{\gamma_0}L\int_{\vec{\sigma}}\sin\theta\left[\vec{\nabla}\theta\times\vec{\nabla}\phi\right]\cdot\vec{\mathrm{d}}\vec{S}.
\end{equation}
However, in general case of non-Euclidean metric the amplitude of the gyrovector can deviate from the value \eqref{eq:G-simple}. 

For angular parameterization the elements of the dissipative tensor \eqref{eq:Dab} are as follows
\begin{equation}\label{eq:D-angles}
\begin{split}
&D_{\alpha\beta}=\frac{M_s}{\gamma_0}\frac{L}{2}\int_{\vec{\sigma}}\Biggl[\partial_\alpha\theta\partial_\beta\theta+\sin^2\theta\Bigl(\partial_\alpha\phi\partial_\beta\phi\\
&+\cos^2\phi\;\partial_\alpha\ln\sqrt{g_{11}}\;\partial_\beta\ln\sqrt{g_{11}}\\
&+\sin^2\phi\;\partial_\alpha\ln\sqrt{g_{22}}\;\partial_\beta\ln\sqrt{g_{22}}\Bigr)\\
&-\sin 2\theta\partial_\alpha\theta\bigl(\cos^2\phi\partial_\beta\ln\sqrt{g_{11}}+\sin^2\phi\partial_\beta\ln\sqrt{g_{22}}\bigr)\\
&-\sin^2\theta\sin 2\phi\partial_\alpha\phi\partial_\beta\ln\sqrt{\frac{g_{22}}{g_{11}}}+(\alpha\leftrightarrow\beta)\Biggr]\mathrm{d}S.
\end{split}
\end{equation}
For an Euclidean metric \eqref{eq:D-angles} is reduced to the familiar formula
\begin{equation}\label{eq:Dab-simple}
D_{\alpha\beta}^\textsc{e}=\frac{M_s}{\gamma_0}L\int_{\vec{\sigma}}\left(\partial_\alpha\theta\partial_\beta\theta+\sin^2\theta\partial_\alpha\phi\partial_\beta\phi\right)\mathrm{d}S
\end{equation}

\subsection{Skyrmion in thin curvilinear shell}\label{app:thiele.skyrm}
Let us assume that we were able to introduce an orthogonal frame of reference $\{\xi^1,\xi^2\}$ on the surface, such that the skyrmion motion can be described by means of the following Ansatz
\begin{subequations}\label{eq:Ansatz}
	\begin{align}
&\theta=\Theta_{\text{pl}}\left(\sqrt{g_{11}(\xi^1-X^1)^2+g_{22}(\xi^2-X^2)^2}\right),\\
&\phi=\arctan\frac{\sqrt{g_{22}}(\xi^2-X^2)}{\sqrt{g_{11}}(\xi^1-X^1)}+\phi_0,
\end{align}
\end{subequations}
where $\Theta_{\text{pl}}(r)$ is skyrmion profile on a plane. Use of the Ansatz \eqref{eq:Ansatz} means that we consider curvature as a small perturbation, which does not change the skyrmion profile. 

In order to obtain the value of the gyrovector $G$, one should substitute \eqref{eq:Ansatz} into \eqref{eq:G-explicit}. If the size of the area of localization of the function $\Theta_{\text{pl}}(r)$ -- the skyrmion radius $r_s$ -- is comparable with typical length scale of change of the metric, then one should expect the deviation of $G$ from its planar value. However, if the skyrmion radius is small, $r_s\ll 1/|\mathcal{H}|$ and $r_s\ll |\mathcal{H}|/|\mathcal{K}|$, then one can assume that $g_{\alpha\alpha}(\xi^1,\xi^2)\approx g_{\alpha\alpha}(X^1,X^2)$ and $\partial_\beta g_{\alpha\alpha}(\xi^1,\xi^2)\approx \partial_\beta g_{\alpha\alpha}(X^1,X^2)$ within the skyrmion core. In this case one can make a change of variables $\sqrt{g_{11}}(\xi^1-X^1)=r\cos\chi$, $\sqrt{g_{22}}(\xi^2-X^2)=r\sin\chi$ in the integral $\int_{\vec{\sigma}}\mathrm{d}S$ and utilize the spatial localization of the function $\sin\Theta_{\text{pl}}$. Now, after the integration over $\chi$ in \eqref{eq:G-int} all terms in \eqref{eq:G-den} but the first one are integrated out. Finally, \eqref{eq:G-general} is reduced to \eqref{eq:G-simple} and one obtains the same result as for the planar case, namely
\begin{equation}\label{eq:G-planar}
G=\frac{M_s}{\gamma_0}L4\pi N_{\text{top}},
\end{equation}
with $N_{\text{top}}$ being topological charge of the \emph{planar} skyrmion.

%\nb[DS]{\alert{@Volodya:} Probably it has a sense to cite \cite{Kravchuk16a} and to define the topological charge using the mapping Jacobian. However it seems that according to \eqref{eq:G-general} an extra factor appears between $G$ and $N_{\text{top}}$ as $\frac{1}{\sqrt{|g(X^1,X^2)|}}$, is it correct?}[21.3.20]

Using the same technique one obtains for the dissipative tensor
\begin{equation}\label{eq:D-skyrm}
D_{\alpha\beta}=\frac{M_s}{\gamma_0}L4\pi\left[C_0g_{\alpha\beta}+C_2\left(\Gamma_{\alpha 1}^1\Gamma_{\beta 1}^1+\Gamma_{\alpha 2}^2\Gamma_{\beta 2}^2\right)\right],
\end{equation}
where $C_0=\frac{1}{4} \int_0^\infty\left[(\partial_r\Theta_{\text{pl}})^2+r^{-2}\sin^2\Theta_{\text{pl}}\right]r\,\mathrm{d}r$, $C_2=\frac{1}{4}\int_0^\infty\sin^2\Theta_{\text{pl}}r\,\mathrm{d}r$ and $\Gamma_{\alpha\beta}^\gamma$ denote the Christoffel symbol of the second kind. Thus, for the surface with the non-Euclidean metric the dissipative tensor is generally non-diagonal even for small radius skyrmion. Note that in the limit case $r_s\to0$ one has $C_0\to1$ and $C_2\to0$.

For the dimensionless time and coordinates introduced in the main text, one writes Thiele equation \eqref{eq:Thiele-crv} in form \eqref{eq:Thiele} if the gyrovector amplitude \eqref{eq:G-planar} is used for $N_{\text{top}}=-1$, and the term with $C_2$ is neglected in the damping tensor \eqref{eq:D-skyrm}. The latter corresponds to the assumption $C_2\ll C_0$ .

Let us estimate the curvature induced corrections to the energy of skyrmion. In order to estimate the exchange energy $E_{\mathrm{ex}}=AL\int\mathscr{E}_{\mathrm{ex}}\mathrm{d}S$ we use the previously derived\cite{Gaididei14} expression for the exchange energy density
\begin{equation}\label{eq:Eex}
\mathscr{E}_{\mathrm{ex}}=\left[\vec{\nabla}\theta-\vec{\Gamma}\right]^2+\left[\sin\theta(\vec{\nabla}\phi-\vec{\Omega})-\cos\theta\partial_\phi\vec{\Gamma}\right]^2.
\end{equation}
Here $\vec{\Gamma}=\vec{g}^\alpha b_{\alpha\beta}m_{||}^\beta(\phi)$. Where $b_{\alpha\beta}$ is the second fundamental form and $m_{||}^\beta=m^\beta(\theta=\pi/2)$ are the magnetization components for the strictly tangential magnetization, namely $m_{||}^1(\phi)=\cos\phi/\sqrt{g_{11}}$ and $m_{||}^2(\phi)=\sin\phi/\sqrt{g_{22}}$. And $\vec{\Omega}$ is the vector of spin connection.\cite{Bowick09,Gaididei14} Using the Ansatz \eqref{eq:Ansatz} and applying the same method as for derivation of the gyrovector, one obtains
\begin{equation}\label{eq:Eex-int}
E_{\mathrm{ex}}=E_{\mathrm{ex}}^0+8\pi AL C_1H\cos\phi_0+\mathcal{O}(H^2,K,|\vec{\Omega}|^2).
\end{equation}
Here $E_{\mathrm{ex}}^0$ is exchange energy of the planar skyrmion, $H$ and $K$ are the mean and Gau{\ss}ian curvature, respectively, and $C_1=\frac14\int\limits_0^\infty \left[ \Theta_{\text{pl}} - \sin\Theta_{\text{pl}} \cos\Theta_{\text{pl}}\right]\mathrm{d}r$. Note that the curvature induced corrections are essentially different for N{\'e}el and Bloch skyrmions: for a Bloch skyrmion ($\phi_0=\pm\pi/2$) the energy \eqref{eq:Eex-int} does not have the contribution linear in curvature. In order to evaluate the quadratic corrections, one should introduce the Riemann normal coordinates \cite{Petrov69,Veblen22} centered on the skyrmion. In this case, the spin connections is determined only by the Gau{\ss}ian curvature of the surface and not by curvature of the curvilinear frame of reference itself. Since we a interested in dynamics of the N{\'e}el skyrmion, we limit ourselves only with the main term in \eqref{eq:Eex-int} with $\cos\phi_0=1$.

Let us now consider the DMI energy $E_\textsc{d}=\mathcal{D}L\int\mathscr{E}_{\textsc{d}}\mathrm{d}S$ with density \cite{Kravchuk16a,Kravchuk18a}
\begin{equation}\label{eq:Edmi-den}
\mathscr{E}_{\textsc{d}}=2\partial_\alpha\theta\, m_{||}^\alpha(\phi)\sin^2\theta-H\cos^2\theta.
\end{equation}
Applying the same procedure as for the exchange energy one obtains
\begin{equation}\label{eq:Edmi-tot}
E_\textsc{d}=E_\textsc{d}^0+8\pi\mathcal{D}LC_2H+\text{const},
\end{equation}
where constant $C_2$ is determined above.

Within the used approximation the anisotropy energy of the skyrmion is a constant independent on the skyrmion position. Collecting now \eqref{eq:Eex-int} and \eqref{eq:Edmi-tot}, and passing to the dimensionless coordinates and curvature $\mathcal{H}=\ell H$ one obtains expression \eqref{eq:E-CV} for the normalized energy.

%\begin{widetext}
%	\listofchanges[title={Comments by DS}]
%\end{widetext}

%
%----------------------------------------------------------------
%

\section{Finite element micromagnetic simulations}\label{app:micromag}

In order to support the theoretical predictions and numerical calculations presented in the manuscript we performed full scale finite element micromagnetic simulations. 
The static equilibrium states of the skyrmions on the Gau{\ss}ian bumps as well as their excitations were obtained by the numerical integration of the Landau-Lifshitz-Gilbert equation using our finite element micromagnetic simulator TetraMag.\cite{Kakay10} The following material parameters were considered to mimic a Pt=Co=AlOx layer structure: $A = 1.6 \times 10^{11}$ J/m being the exchange constant, $\mu_0 M_s = 1.38$ T the saturation magnetization and $K_u = 1.3 \times 10^6$ J/$\text{m}^3$ for the uniaxial anisotropy constant pointing along the surface normal. The effective anisotropy constant for the renormalised magnetostatic case was set to $K_{eff} = K_u  - 2 \pi M_s^2 = 5.1 \times 10^5$ J/$\text{m}^3$. The magnetic length is therefore $\ell=\sqrt{A/K_{eff}} =5.6$ nm. 

To study the skyrmion excitations a Gau{\ss}ian bump defined by equation (4) with a thickness of 1 nm and 200 nm in diameter was used. The bump amplitude and radius was varies in the simulations. The static skyrmion profiles were calculated with a Gilbert damping $\alpha = 0.5$, while the skyrmion dynamics is simulated with  $\alpha = 0.02$. The enrgy contribution from the intrinsic DMI is implemented into our micromagnetic simulator TetraMag in the following form: $ \mathcal{E}_{DMI} = m_{\bm{n}}\bm{\nabla}\cdot\bm{m} - \bm{m}\cdot\bm{\nabla} m_{\bm{n}}$ with the corresponding effective field $ \bm{H}_{DMI} = - \dfrac{2 \mathcal{D}}{M_S} \left[\bm{n}\bm{\nabla}\cdot\bm{m}- \bm{\nabla}\left(\bm{m}\cdot\bm{n} \right) \right] $, where  $\bm{n}$ is the surface normal vector, $\bm{m}=M/M_s$ the unit magnetization vector, $m_{\bm{n}}$ the surface normal component of the magnetization vector field and $\mathcal{D} = 2.8\times10^{-3} \;\mathrm{J/m^2}\; (d = 0.98)$ being the DMI constant.

The following fields have been used to excite thye different magnon modes of the skyrmion localised to the center of the Gau{\ss}ian bump:

 - the skyrmion \textit{gyromotion} (corresponding to the mode with the azimuthal quantum number $\mu = 1$) is excited by applying an inhomogeneous external magnetic field $\bm{B} = \hat{\bm{z}}B_0 x\mathrm{H}(L-\sqrt{x^2 + y^2})/L$, $L=50$ nm and $B_0$ = 50~mT; the $\mathrm{H}(\bullet)$ is the Heaviside step function. 
 
 - the \textit{breathing mode} ($\mu = 0$) is excited with a uniform magnetic field, $\bm{B} = \hat{\bm{z}}B_0$ applied for a time duration of about 150 ps and field strength of $B_0$ = 50 mT. 
 
 - the \textit{CCW gyrotropic mode} ($\mu = -1$) was excited with a uniform field $\bm{B} =\hat{\bm{x}} B_0 $ applied for 200 ps and field strength of $B_0$ = 50 mT. 
 
 - the \textit{elliptical mode} ($\mu = 2$) was excited by an inhomogeneous external magnetic field pulse $\bm{B} =\hat{\bm{z}} B_0  \cos(2\chi)\mathrm{H}(L-\sqrt{x^2 + y^2})$ applied for a time duration of 200 ps and field strength of $B_0$ = 50 mT. 
 
 - to achieve the excitation of modes with \textit{azimuthal mode} number $\mu = 3$ we applied the following field pulse $\bm{B} = \hat{\bm{z}}B_0  \cos(3\chi)\mathrm{H}(L-\sqrt{x^2 + y^2})\sqrt{x^2 + y^2}/L$ for a time period of 50 ps to 100 ps, depending on the bump amplitude and the fields strength was set to $B_0$ = 50 mT. 
 
 In order to determine the mode frequencies we performed a Fast Fourier Transform of the magnetization of every single discretization node over a relatively long time interval (minimum 10 periods of simulations). The power spectra obtained from the eigenvalue problem (equation (\ref{eq:EVP})) and with the collective variables approach (\ref{eq:omega1}) are in perfect agreement with that calculated with the full-scale finite element micromagnetic simulations.

%\bibliography{soliton}
%apsrev4-2.bst 2019-01-14 (MD) hand-edited version of apsrev4-1.bst
%Control: key (0)
%Control: author (8) initials jnrlst
%Control: editor formatted (1) identically to author
%Control: production of article title (0) allowed
%Control: page (0) single
%Control: year (1) truncated
%Control: production of eprint (0) enabled
%

\end{document}